 \documentclass[pmlr,twocolumn,10pt]{jmlr} 

\usepackage{booktabs}
\usepackage{adjustbox}
\usepackage{multirow}
\usepackage{siunitx}

\theorembodyfont{\upshape}
\theoremheaderfont{\scshape}
\theorempostheader{:}
\theoremsep{\newline}

\jmlrvolume{225}
\jmlryear{2023}
\jmlrsubmitted{LEAVE UNSET}
\jmlrpublished{LEAVE UNSET}
\jmlrworkshop{Machine Learning for Health (ML4H) 2023} 

 \title[
Automated Cardiovascular Record Retrieval by Multimodal Learning
]{
Automated Cardiovascular Record Retrieval by Multimodal Learning between Electrocardiogram and Clinical Report
}

\author{\Name{Jielin Qiu$^{*1}$} \Email{jielinq@cs.cmu.edu}  \\
\Name{Jiacheng Zhu$^{*2}$} \Email{zjc@mit.edu} \\
\Name{Shiqi Liu$^1$} \Email{shiqiliu@andrew.cmu.edu} \\
\Name{William Han$^1$} \Email{wjhan@andrew.cmu.edu} \\
\Name{Jingqi Zhang$^1$} \Email{jingqiz@andrew.cmu.edu} \\
\Name{Chaojing Duan$^3$} \Email{chaojind@andrew.cmu.edu} \\
\Name{Michael A. Rosenberg$^4$} \Email{michael.a.rosenberg@cuanschutz.edu} \\
\Name{Emerson Liu$^3$} \Email{emersonliu@msn.com} \\
\Name{Douglas Weber$^1$} \Email{dougweber@cmu.edu} \\
\Name{Ding Zhao$^1$} \Email{dingzhao@cmu.edu} \\
\addr $^1$ Carnegie Mellon University\\
$^2$ MIT CSAIL \\
$^3$ Allegheny Health Network \\
$^4$ University of Colorado Denver
\AND
\footnotemark[1] \addr {\normalfont \footnotesize marked as equal contribution}
}

\begin{document}

\maketitle

\begin{abstract}
Automated interpretation of electrocardiograms (ECG) has garnered significant attention with the advancements in machine learning methodologies. Despite the growing interest, most current studies focus solely on classification or regression tasks which overlook a crucial aspect of clinical cardio-disease diagnosis: the diagnostic report generated by experienced human clinicians. In this paper, we introduce a novel approach to ECG interpretation, leveraging recent breakthroughs in Large Language Models (LLMs) and Vision-Transformer (ViT) models. Rather than treating ECG diagnosis as a classification or regression task, we propose an alternative method of automatically identifying the most similar clinical cases based on the input ECG data. Also, since interpreting ECG as images is more affordable and accessible, we process ECG as encoded images and adopt a vision-language learning paradigm to jointly learn vision-language alignment between encoded ECG images and ECG diagnosis reports. Encoding ECG into images can result in an efficient ECG retrieval system, which will be highly practical and useful in clinical applications. 
More importantly, our findings could serve as a crucial resource for providing diagnostic services in underdevelopment regions.
\end{abstract}
\begin{keywords}
Cardiovascular, Retrieval, Multimodal, Electrocardiogram
\end{keywords}

\section{Introduction}

Cardiovascular diseases, such as heart attacks and strokes, account for the majority of global deaths. ECG is a vital tool in cardiology and electrophysiology, providing valuable information about the heart's structure, electrical activity, and potential systemic conditions through waveform changes in timing and morphology. Accurate interpretation of clinical ECGs is critical, as it remains a primary method for identifying cardiac abnormalities and screening populations at risk of heart-related issues.

\begin{figure*}[t]
\centering
\includegraphics[width=0.75\textwidth]{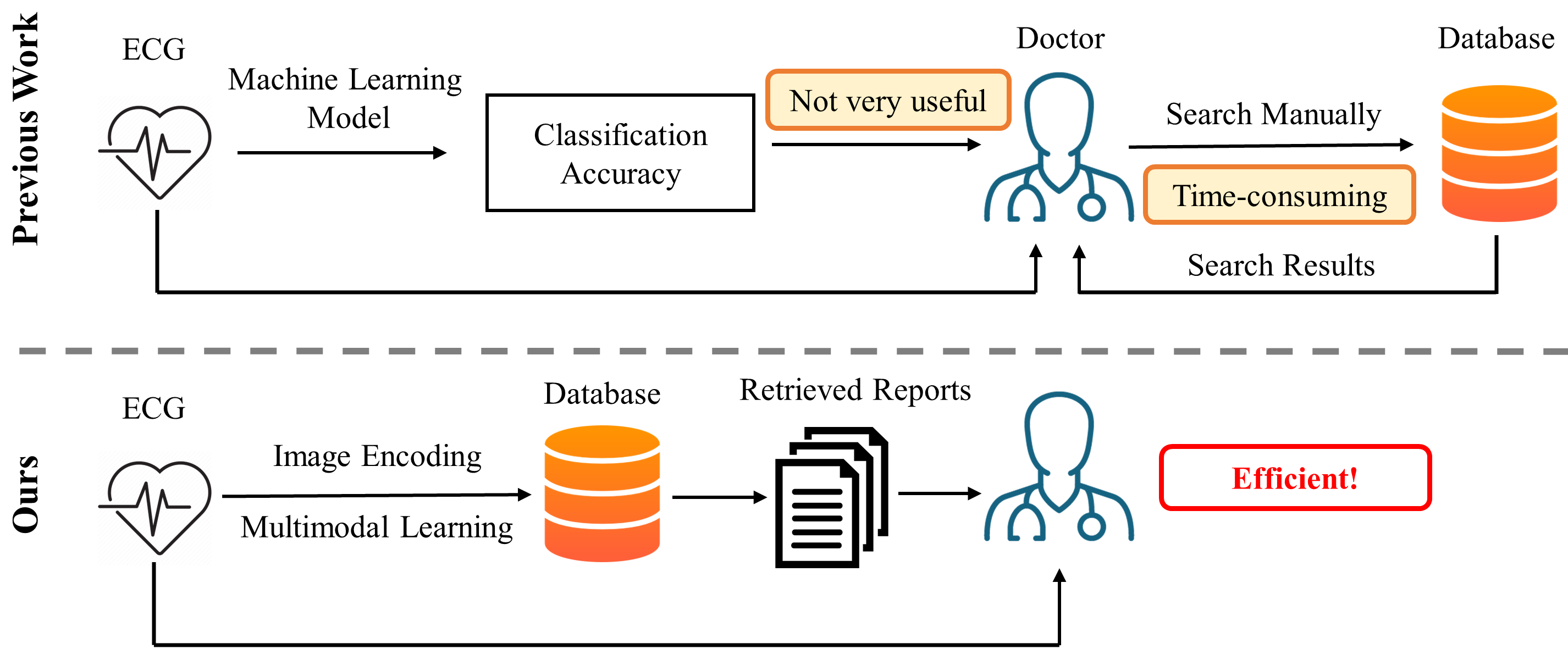}
\caption{\small{Prior studies only provided disease prediction accuracy for machine learning models, leaving doctors to cross-reference databases for precise diagnoses. In contrast, our method retrieves relevant past studies in the database, greatly reducing doctors' manual search efforts and improving patient care efficiency.}}
\label{fig:model}
\vspace{-20pt}
\end{figure*}

The precise interpretation of ECGs is essential
for providing timely, efficient, and cost-effective interventions for acute cardiac conditions.
Machine learning (ML) algorithms have been used to assist ECG interpretation, including disease classification~\citep{pmlr-v149-nonaka21a,Khurshid2021ElectrocardiogrambasedDL,Raghunath2021DeepNN,Giudicessi2021ArtificialIA,Strodthoff2021DeepLF}, adversarial attack~\citep{han2020deep,hossain2021_ecg_adv_gan,chen2020_aaai_ecg_adv}, data augmentation~\citep{raghu22a,nonaka2020electrocardiogram_ecg_data_aug}, contrastive learning~\citep{gopal20213kg}, and the application of transformer models~\citep{Che2021ConstrainedTN,Natarajan2020AWA,Behinaein2021ATA}. 

Most current machine learning ECG interpretation frameworks have practical limitations. They mainly use the ECG signal as input and diagnosis as a label, adapting to supervised learning as in other fields. However, ECG diagnosis is multifaceted, involving a complex hierarchy of disorders. For instance, the ST/T changes class can be divided into subclasses like ischemic in anterior leads (ISCA), ischemic in inferior leads (ISCI), non-specific ischemic (ISC), and non-specific ST changes (NST)~\citep{guglin2006common_ecg}.
In practice, physicians commonly provide detailed ECG reports~\citep{Wagner2020PTBXLAL} containing nuanced signal features and categorical diagnoses.
AI-powered ECG frameworks, however, assume digital ECG processing via advanced systems. In reality, paper-printed ECGs~\citep{zhang2023artificial_ecg_image} from ECG monitor machines~\citep{olson2013mayo} are predominantly used by patients and doctors. Notably, paper-printed ECGs are the sole protocols in \textbf{underdeveloped regions}.

To overcome the aforementioned limitations, our goal is to enhance ECG interpretation automation by addressing the following challenge: \underline{\textit{Can we automatically match it with the most similar}}  \underline{\textit{ ECG records in the database?}}? This involves leveraging joint inference between the ECG signal and expert-written reports. This functionality can greatly aid in diagnosing common diseases like arrhythmia~\citep{hong2020cardiolearn,fu2021artificial}, reducing physicians' workload~\citep{hannun2019cardiologist}.
Additionally, this ECG data retrieval system can assist in diagnosing complex conditions such as atrial fibrillation and contractile dysfunction~\citep{attia2019artificial,attia2019screening}, which pose challenges for supervised learning networks due to limited training data.


To realize this goal, we introduce an ECG-Text retrieval system that employs a multimodal information retrieval framework to automatically fetch expert-written reports along with corresponding ECG records. From a pragmatic standpoint, we treat ECG data as image input and employ various featurization methods. Our model is designed to discern the similarity score between these two modalities, enabling automatic identification of correspondences between ECG images and human language descriptions. Our contributions are outlined as follows:
\begin{itemize}
    \vspace{-5pt}
    \item Our approach aims to improve ML-driven ECG automatic diagnosis by tackling the multimodal retrieval challenge and training to align the two modalities.
    \vspace{-5pt}
    \item  We present a robust framework that provides clinicians with a practical and efficient method to automatically search and identify similar ECG records for newly acquired ECG data.
    \vspace{-5pt}
    \item Building upon progress in image-text alignment research, we highlight the treatment of ECG signals as images and introduce diverse preprocessing methods. This strategy makes our approach practical and readily applicable, given the widespread use of commercial ECG machines.
    \vspace{-5pt}
\end{itemize}

\section{Related Work}

\paragraph{Multimodal Learning in Healthcare}

The computational field of machine learning has faced the multimodal nature of clinical expert decision-making. \citet{Kline2022MultimodalML} summarized the current studies in multimodal learning in healthcare applications and identified topics ripe for future research. \citet{Amal2022UseOM} reviewed multimodal data fusion and machine learning in cardiovascular medicine. For example, the detection of cardiac amyloidosis can benefit from fusing ECG signals and echocardiograms with convolution neural networks~\citep{goto2021artificial}. 
The multimodal approach also helps as combining salient physiological signals and EHR data can effectively predict the onset of hemodynamic decompensation~\citep{hernandez2021multimodal}. However, our study is the first to investigate the multimodal properties between ECG and natural language data.

\paragraph{Encode Time Series Signals into Images}  
Deep learning has been successfully applied to automate ECG diagnosis~\citep{han2020deep_adv_ecg}.
These methods are usually based on raw ECG signal data and corresponding features \citep{Kiranyaz2015ConvolutionalNN,Zhu2022GeoECGDA}. However, traditionally, ECG data is transformed into printed images with waveforms and interpreted by trained clinicians~\citep{sangha2022automated_ecg}.
To harness recent advancements in deep learning and computer vision, making ECG interpretation more practical and accessible, machine learning approaches that treat ECG data as image features have been investigated. An early approach combined either ECG images or signals~\citep{sangha2022automated_ecg} as inputs for cardiac disease diagnosis by a convolutional neural network based on the EfficientNet architecture. The idea of interpreting printed ECG papers has also been shown to be effective for diagnosing left ventricular (LV) systolic dysfunction~\citep{sangha2022detection_LV}. Additionally, digitizing printed ECG papers by scanning and processing raw printed images~\citep{wu2022fully_digit_ecg} is a critical task. Similarly, an automated ECG diagnostic pipeline employing paper-ECG images can facilitate accessible diagnostic services in regions with limited healthcare information systems.

\vspace{-5pt}
\section{Methods}



Our approach comprises two key components: (1) the conversion of ECG time series signals into images, and (2) the utilization of these encoded ECG images and their corresponding clinical reports to construct an ECG record retrieval system. We delve into the specifics in Section~\ref{sec:encoding} and Section~\ref{sec:retrieval-system} respectively.

\subsection{Encode ECG Signals into Images}\label{sec:encoding}

In our study, we employed three different encoding methods to convert ECG time series signals into visual formats: Markov Transition Field (MTF), Gramian Angular Field (GAF), and Recurrence Plot (RP). Each technique's detailed explanation is provided in the subsequent sections, with further particulars available in Appendix~\ref{sec:appendix-encoding} due to space limit.

\begin{figure}[t]
\centering
\includegraphics[width=0.49\textwidth]{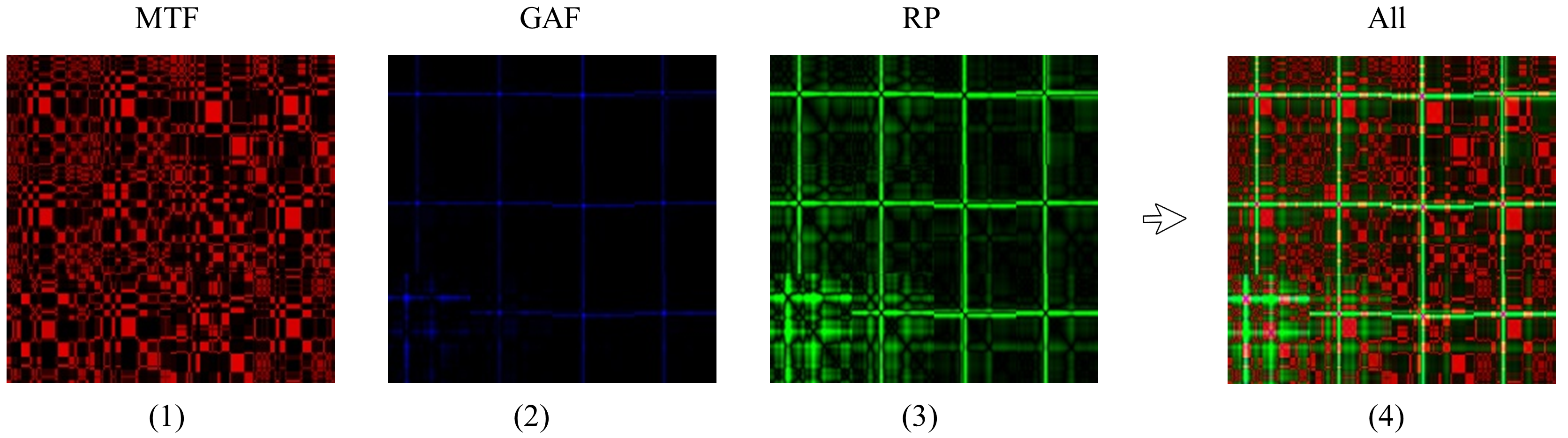}
\caption{\small{Examples of encoded ECG images by (1) MTF; (2) GAF; (3) RP; and (4) combine all three methods in three channels.}}
\vspace{-20pt}
\label{fig:example}
\end{figure}

\subsubsection{Markov Transition Field (MTF)}

Markov Transition Field (MTF) is a method of transforming time series data, such as ECG signals, into visual representations. MTF works by calculating transition probabilities between adjacent data points in a time series, and then using these probabilities to generate a matrix of color-coded pixels.
Given a ECG time series $X$, the $Q$ quantile bins are identified, and each data point $x_i$ is assigned to its corresponding bin $q_j (j\in [1,Q])$. The resulting weighted adjacency matrix $W$, constructed using a first-order Markov chain model along the time axis, reflects the transition probabilities among the quantile bins. The frequency with which a data point in quantile bin $q_j$ is followed by a point in bin $q_i$ determines the value of the corresponding entry $w_{i,j}$ in $W$. Although $W$ represents the Markov transition matrix after normalization by $\sum_{j} { w_{ij}} = 1$, it is insensitive to the distribution of $X$ and the temporal dependencies between time steps $t_i$, resulting in a loss of information. To address this issue, the Markov Transition Field (MTF) $M$ is defined as follows:

\vspace{-5pt}
\begin{equation}\scriptsize
\begin{bmatrix}
w_{ij|{x_1 \in q_i},{x_1 \in q_j}}&w_{ij|{x_1 \in q_i},{x_2 \in q_j}}    & \cdots\ &w_{ij|{x_1 \in q_i},{x_n \in q_j}}\\
w_{ij|{x_2 \in q_i},{x_1 \in q_j}}&w_{ij|{x_2 \in q_i},{x_2 \in q_j}}    & \cdots\ & w_{ij|{x_2 \in q_i},{x_n \in q_j}}\\
\vdots &\vdots   & \ddots  & \vdots  \\
w_{ij|{x_n \in q_i},{x_1 \in q_j}} &w_{ij|{x_n \in q_i},{x_2 \in q_j}}  & \cdots\ & w_{ij|{x_n \in q_i},{x_n \in q_j}}\\
\end{bmatrix}
\end{equation}
It involves building a $Q \times Q$ Markov transition matrix $W$ by dividing the time series data into $Q$ quantile bins, where $q_i$ and $q_j (q\in [1,Q])$ represent the quantile bins that contain the data at time stamps $i$ and $j$ along the temporal axis. The MTF matrix $M$ encodes the transition probabilities of the time series by spreading out the transition probability values from matrix $W$ along the magnitude axis to $M$ while taking into consideration the temporal positions. At each pixel $M_{ij}$, the probability of transitioning from the quantile at time step $i$ to the quantile at time step $j$ is assigned. In this way, the MTF matrix $M$ captures the multi-span transition probabilities of the time series. The entry $M_{i,j||i-j|=k }$ in $M$ represents the transition probability between points with a time interval of $k$, where $M_{i,j||j-i}=1$ represents the transition process along the time axis with a skip step. The main diagonal $M_{ii}$ in $M$ is a special case when $k = 0$ and captures the probability of transitioning from each quantile to itself, i.e., the self-transition probability, at time step $i$.

\subsubsection{Gramian Angular Field (GAF)}

Gramian Angular Field (GAF) \citep{Wang2014EncodingTS} is another method for transforming ECG time series signals into visual representations. GAF generates a matrix of cosine and sine values based on the pairwise differences between the original data points in the time series. This matrix is then transformed into an image, where each pixel corresponds to a particular combination of cosine and sine values. Similar to MTF, the resulting image captures important features of the original ECG signal, such as patterns and trends, which can aid in the interpretation and analysis of the data. 

The Gramian Angular Field (GAF) \citep{Wang2014EncodingTS} method represents time series data in a polar coordinate system instead of using the traditional Cartesian coordinates. In the Gramian matrix of GAF, each element corresponds to the cosine of the summation of angles.
The rescaled time series $\tilde{X}$ of $n$ real-valued observations are transformed to fall within the range of [$-1,1$] or [$0,1$] using the formula:
\begin{equation}\scriptsize
\tilde{x}^i_{-1}=\frac{(x_i -max(X) +(x_i -min(X))}{max(X)-min(X)}
\end{equation}
\begin{equation}\scriptsize
or \quad \tilde{x}^i_{0}=\frac{x_i -min(X)}{max(X)-min(X)}
\end{equation}
Then, by encoding the value as the angular cosine and the time stamp as the radius, we represent the rescaled time series $\tilde{X}$ in polar coordinates as follows:
\begin{equation}\scriptsize
\phi=arccos(\tilde{x}_i), \quad -1 \le \tilde{x}_i \le 1, \quad \tilde{x}_i \in \tilde{X}, \quad r=\frac{t_i}{N}, \quad t_i \in N
\end{equation}
Here, $t_i$ is the time stamp, and $N$ is a constant factor that regulates the span of the polar coordinate system. This encoding technique is a novel way to visualize time series data, where the values transform among different angular positions on the spanning circles as time passes, resembling water rippling. The encoding map is bijective, and it preserves absolute temporal relations, unlike Cartesian coordinates. The angular cosine function is monotonic for $\phi \in [0,\pi]$, producing a unique result in the polar coordinate system with a one-to-one inverse map.


We utilize the angular perspective of the polar coordinate system to examine temporal correlations between different time intervals by calculating the trigonometric sum/difference between each point. Specifically, we define the Gramian Summation Angular Field (GASF) and Gramian Difference Angular Field (GADF) as follows:
\begin{equation}\scriptsize
\begin{aligned}
GASF=[cos({\phi}_i + {\phi}_j)]=\tilde{X}' \cdot \tilde{X} - \sqrt{I-\tilde{X}^2}' \cdot \sqrt{I-\tilde{X}^2}
\end{aligned}
\end{equation}
\begin{equation}\scriptsize
\begin{aligned}
GADF=[sin({\phi}_i - {\phi}_j)]=\sqrt{I-\tilde{X}^2}' \cdot \tilde{X} - \tilde{X}' \cdot \sqrt{I-\tilde{X}^2}
\end{aligned}
\end{equation}
Here, $I$ is the unit row vector $[1,1,...,1]$. After transforming the time series into the polar coordinate system, we treat each time step as a 1-D metric space. Defining the inner product as follows:
\begin{equation}\scriptsize
\begin{aligned}
<x,y >_1= x\cdot y - \sqrt{1 - x^2 } \cdot \sqrt{1 - y^2}
\end{aligned}
\end{equation}
\begin{equation}\scriptsize
\begin{aligned}
< x,y >_2=\sqrt{1 - x^2 } \cdot y - x \cdot \sqrt{1 - y^2}
\end{aligned}
\end{equation}
The two types of Gramian Angular Fields (GAFs) are actually quasi-Gramian matrices $[< \tilde{x_1}, \tilde{x_1} >]$.

The Gramian Angular Fields (GAFs) offer multiple benefits. First, they enable the retention of temporal relationships, as the position's movement from the top-left to the bottom-right corresponds to the increase in time. The GAFs incorporate temporal correlations since $G_{i,j||i-j|=k}$ symbolizes the relative correlation due to the superimposition/difference of directions concerning time interval $k$. The main diagonal $G_{i,i}$ is a special case for $k=0$, containing the original angular/value information.

\subsubsection{Recurrence Plot (RP)}

Recurrence Plot (RP) \citep{Eckmann1987RecurrencePO} is a non-linear time series analysis technique that can also be applied to transform ECG time series signals into visual representations. RP generates a square matrix that reflects the similarity between all pairs of data points in the time series. The matrix is constructed by measuring the distance between each pair of data points and comparing them to a predefined threshold value. RP has been shown to be effective in capturing complex patterns in ECG signals, such as P-waves and QRS complexes, which are important for the accurate diagnosis of cardiovascular diseases. 

Given a time series $(x_1, \ldots, x_n)$, we can extract trajectories from it as follows:
\begin{equation}\scriptsize
\begin{aligned}
\vec{x}i = (x_i, x{i + \tau}, \ldots, x_{i + (m - 1)\tau}), \quad
\forall i \in {1, \ldots, n - (m - 1)\tau }
\end{aligned}
\end{equation}
Here, $m$ denotes the dimension of the trajectories, and $\tau$ is the time delay. Once we have extracted the trajectories, we can create a recurrence plot, denoted by $R$, which is essentially the pairwise distance between the trajectories. Formally, we define $R_{i, j}$ as:
\begin{equation}\scriptsize
\begin{aligned}
R_{i, j} = \Theta(\varepsilon - | \vec{x}_i - \vec{x}_j |), \quad
\forall i,j \in {1, \ldots, n - (m - 1)\tau }
\end{aligned}
\end{equation}
Here, $\Theta$ is the Heaviside step function, and $\varepsilon$ is the threshold. The recurrence plot helps us visualize the structure and patterns of the time series by preserving the temporal dependencies and revealing the relative correlations between the extracted trajectories.

\subsection{Retrieval System}\label{sec:retrieval-system}

This section commences with an overview of the model architecture, followed by a detailed account of the training objectives.  An elaborate illustration of the model architecture is presented in Figure~\ref{fig:model}. Our model follows \citet{ALBEF,Li2022BLIPBL}, which includes a vision encoder responsible for processing visual information, a language encoder dedicated to understanding textual data, and a multimodal encoder that integrates information from both the vision and language encoders to form a robust representation.

\begin{figure*}[t]
\centering
\includegraphics[width=0.75\textwidth]{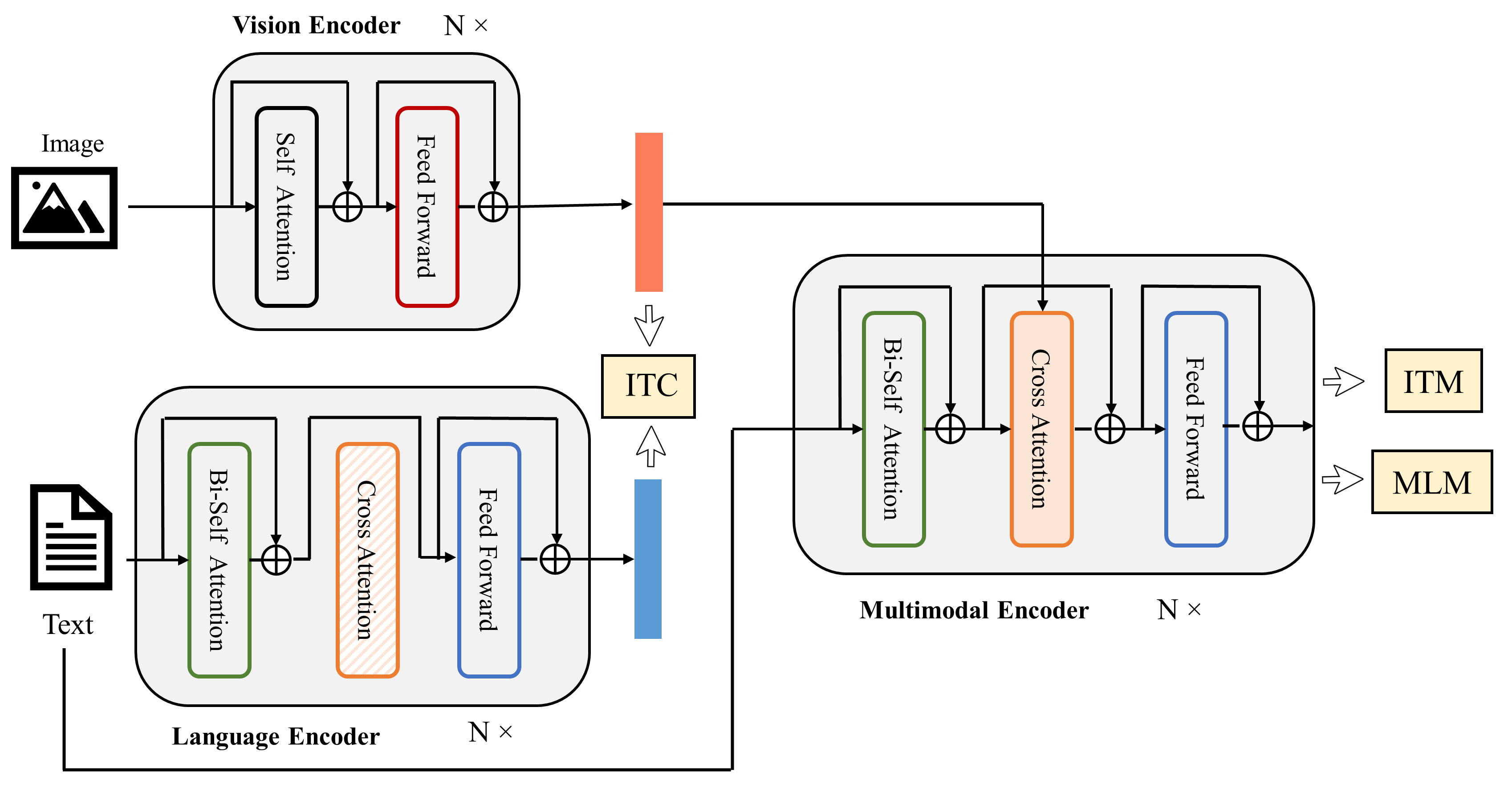}
\caption{\small{The overall architecture of our model, which comprises a vision encoder responsible for processing visual data, a language encoder that focuses on comprehending textual information, and a multimodal encoder that combines the input from both the vision and language encoders to fuse comprehensive representations.}}
\vspace{-20pt}
\label{fig:model}
\end{figure*}

\paragraph{Vision Encoder} Our current vision encoder architecture is based on a visual transformer~\citep{vit}, which implements a patch-based processing approach that encodes an input image into a sequence of embeddings. This is achieved by dividing the image into patches and then performing a sequence of encoding operations on each patch. In addition, an extra \texttt{[CLS]} token is included to represent the global image feature. This approach has been shown to be more computation-friendly than using pre-trained object detectors for visual feature extraction~\citep{uniter} and has been adopted by more recent methods such as ALBEF and ViLT~\citep{ALBEF,ViLT}.
Specifically, given an input image $I$, the ViT-based vision encoder generates a sequence of embeddings: ${\Vec{v}_\mathrm{cls}, \Vec{v}_1,...,\Vec{v}N}$. Here, $\Vec{v}\mathrm{cls}$ represents the embedding of the \texttt{[CLS]} token, and the remaining $\Vec{v}_i$ represents the patch embeddings. It is worth noting that this patch-based processing approach allows the vision encoder to capture fine-grained details of the input image, which can be important for downstream tasks that require a high level of visual understanding.

\paragraph{Language Encoder} Our text encoder is based on the highly effective BERT architecture~\citep{bert}, which employs a \texttt{[CLS]} token appended to the beginning of the input text to provide a summary of the sentence. The encoder also utilizes a bi-directional self-attention mechanism to generate representations for each of the input tokens. This approach is highly effective for capturing the context and meaning of each token in the input text, enabling the model to better understand the overall meaning of the text.
When processing an input text $T$, the text encoder generates a sequence of embeddings ${\Vec{w}_\mathrm{cls}, \Vec{w}_1,...,\Vec{w}_N}$, where $\Vec{w}_\mathrm{cls}$ represents the embedding of the \texttt{[CLS]} token, and the remaining $\Vec{w}_i$ represent the embeddings of the individual input tokens. This sequence of embeddings is then passed to the multimodal encoder to be combined with the visual embeddings generated by the vision encoder.

\paragraph{Multimodal Encoder} The multimodal encoder is a complex module that plays a critical role in enabling the model to learn the relationships between the visual and textual inputs. To achieve this, it incorporates an additional cross-attention (CA) layer that sits between the self-attention (SA) layer and the feed-forward network (FFN) for each transformer block of the text encoder. By doing so, the model can attend to both the textual and visual inputs and build better representations of the image-text pair.
To create a multimodal representation of the image-text pair, the text input is modified by appending a task-specific \texttt{[Encode]} token at the end of the sequence, which is then fed into the multimodal encoder. The output embedding of this token is used as the final representation of the image-text pair. The embedding layers, CA layers, and FFN share similar functionality between encoding and decoding tasks, which means that they can be shared to improve training efficiency and benefit from multi-task learning.
Additionally, the cross-attention layer introduces another set of attention weights to the model, which requires additional computation and increases the number of parameters to be learned. However, this additional complexity is necessary to enable the model to learn the relationships between the visual and textual inputs and to achieve state-of-the-art performance on various image-text tasks.

\subsection{Loss Objectives}

There are three objectives during learning, including Image-Text Contrastive (ITC) Loss, Image-Text Matching  (ITM) Loss, and Mask Language Modeling (MLM) Loss.  An overview of each loss is provided in the subsequent sections. More details can be found in the Appendix due to the page limit.

\paragraph{Image-Text Contrastive Loss (ITC)}  To compute the ITC loss, we follow the approach proposed by~\citet{ALBEF}, which introduces a momentum encoder to generate features and creates soft labels from the momentum encoder to serve as training targets. The soft labels help account for the potential positive samples in the negative pairs and improve the quality of the learned representations. The model learns a similarity function represented by 
$s=g_v(\Vec{v}_\mathrm{cls})^\top g_w(\Vec{w}_\mathrm{cls})$, 
which aims to increase the similarity scores for matching image-text pairs. Here, $g_v$ and $g_w$ refer to linear transformations that convert the \texttt{[CLS]} embeddings into lower-dimensional, normalized (256-d) representations. Following the MoCo approach~\citep{moco}, we use two queues to store the most recent $M$ image-text representations obtained from the momentum unimodal encoders. The features obtained from the momentum encoders are normalized and denoted by $g'v(\Vec{v}'_\mathrm{cls})$ and $g'w(\Vec{w}'_\mathrm{cls})$. To calculate the similarity score between an image-text pair and a text-image pair, we define $s(I,T) = g_v(\Vec{v}_\mathrm{cls})^\top g'w(\Vec{w}'_\mathrm{cls})$ and $s(T,I) = g_w(\Vec{w}_\mathrm{cls})^\top g'v(\Vec{v}'_\mathrm{cls})$, respectively.
We use the softmax-normalized image-to-text and text-to-image similarity to calculate each image and text. This is represented by the equations below, where $\tau$ is a temperature parameter that can be learned:
\begin{equation}\scriptsize
\label{eqn:sim}
p_m^\mathrm{i2t}(I) = \frac{\exp (s(I,T_m) / \tau)}{\sum_{m=1}^M \exp (s(I,T_m)/ \tau)},
\end{equation}
\begin{equation}\scriptsize
p_m^\mathrm{t2i}(T) = \frac{\exp (s(T,I_m)/ \tau)}{\sum_{m=1}^M \exp (s(T,I_m)/ \tau)}
\end{equation}
We represent the ground-truth one-hot similarity as $\Vec{y}^\mathrm{i2t}(I)$ and $\Vec{y}^\mathrm{t2i}(T)$, where negative pairs have a probability of 0, and the positive pair has a probability of 1. The image-text contrastive loss is defined as the cross-entropy $\mathrm{H}$ between $\Vec{p}$ and $\Vec{y}$, which is shown in the following equation:
\begin{equation}\scriptsize
\label{eqn:itc}
\mathcal{L}\mathrm{itc} = \frac{1}{2} \mathbb{E}{(I,T)\sim D} \big[ \mathrm{H}(\Vec{y}^\mathrm{i2t}(I),\Vec{p}^\mathrm{i2t}(I)) + \mathrm{H}(\Vec{y}^\mathrm{t2i}(T),\Vec{p}^\mathrm{t2i}(T)) \big]
\end{equation}

\paragraph{Image-Text Matching Loss (ITM)} ITM is a binary classification task where the model predicts whether an image-text pair is positive (matched) or negative (unmatched) based on its multimodal feature. The ITM head, which is a linear layer, is used to make this prediction.
To obtain the joint representation of the image-text pair, we use the output embedding of the \texttt{[CLS]} token from the multimodal encoder, and then append a fully-connected (FC) layer followed by softmax to predict a two-class probability $p^\mathrm{itm}$. The ITM loss is defined as:
$\mathcal{L}\mathrm{itm} = \mathbb{E}{(I,T)\sim D} \mathrm{H} (\Vec{y}^\textrm{itm}, \Vec{p}^\textrm{itm}(I,T))$,
where $\Vec{y}^\textrm{itm}$ is a 2-dimensional one-hot vector representing the ground-truth label.
To improve the selection of negative pairs, we employ a strategy called hard negative mining, as proposed by~\citet{ALBEF}. This strategy involves selecting negative pairs that have a higher contrastive similarity within a batch.

\paragraph{Mask Language Modeling Loss (MLM)} The Mask Language Modeling Loss (MLM) is used to predict masked words using both the image and contextual text. In this loss, we randomly mask out input tokens with a probability of 15\% and replace them with the special token \texttt{[MASK]}, with 10\% random tokens, 10\% unchanged, and 80\% \texttt{[MASK]} replacements following the BERT approach. The predicted probability of a masked token is denoted by $\Vec{p}^\textrm{msk}(I,\hat{T})$, where $\hat{T}$ represents the masked text. The cross-entropy loss is used to minimize the difference between the predicted and ground-truth distributions, which is expressed as follows:
\begin{equation}\scriptsize
\label{eqn:mlm}
\mathcal{L}\mathrm{mlm} = \mathbb{E}{(I,\hat{T})\sim D} \mathrm{H} (\Vec{y}^\textrm{msk}, \Vec{p}^\textrm{msk}(I,\hat{T}))
\end{equation}
$\Vec{y}^\textrm{msk}$ represents a one-hot vocabulary distribution and the ground-truth token has a probability of 1.

\section{Experiments}

\subsection{Dataset and Preprocessing}

Our experiments were conducted using the PTB-XL dataset \citep{Wagner2020PTBXLAL}, which comprises clinical 12-lead ECG signals that are 10 seconds in length. The dataset includes five different conditions: Normal ECG (NORM), Myocardial Infarction (MI), ST/T Change (STTC), Conduction Disturbance (CD), and Hypertrophy (HYP). The waveform files are stored in the WaveForm DataBase (WFDB) format and have a precision of 16 bits at a resolution of 1$\mu$V/LSB, with a sampling frequency of 100Hz.
The raw waveform data was annotated by up to two cardiologists who assigned one or more ECG statements to each record, resulting in a total of 71 different ECG statements that conform to the SCP-ECG standard. These statements cover diagnostic, form, and rhythm-related information. Additionally, the dataset contains extensive metadata on demographics, infarction characteristics, likelihoods for diagnostic ECG statements, as well as annotated signal properties.
To convert the time series data into a spectrum, we leveraged the WFDB library \citep{wfdb} to read the raw data and performed Fast Fourier Transform (FFT). In order to eliminate noise, we implemented n-points window filtering, and to eliminate power frequency interference, which occurs at 50Hz, we employed notch processing with a quality factor of 30 \citep{Qiu2022CardiacDD}.

\subsection{Experimental Setting}

The initialization of our visual encoder, text encoder, and multimodal encoder was carried out using the image encoder, text encoder, and image-grounded text encoder from \citet{Li2022BLIPBL}, respectively. Specifically, the visual encoder was based on ViT \citep{Dosovitskiy2020AnII} pre-trained on ImageNet \citep{He2015DelvingDI}, while the other two encoders were initialized from BERT \citep{Devlin2019BERTPO}. All three encoders were trained on a dataset consisting of 14M images from COCO \citep{Lin2014MicrosoftCC}, Visual Genome \citep{Krishna2016VisualGC}, Conceptual Captions \citep{Sharma2018ConceptualCA}, Conceptual 12M \citep{Changpinyo2021Conceptual1P}, and SBU captions \citep{Ordonez2011Im2TextDI}, as described in \citet{Li2022BLIPBL}.

Next, we fine-tuned the three encoders on our ECG image data using the AdamW optimizer \citep{Loshchilov2017DecoupledWD} with a weight decay of 0.05. The learning rate was warmed-up to 3e-4 (for ViT-B) / 2e-4 (for ViT-L) and decayed linearly with a rate of 0.85. During fine-tuning, we randomly cropped images to a resolution of 384 × 384. Our experiments were conducted on 4 NVIDIA A6000. We evaluate our models using the recall at K (R@K) metric, where K = ${1, 5, 10}$, and report the RSUM, which is the sum of the recall metrics at K = ${1, 5, 10}$ for both image and text retrieval tasks.

\subsection{Experimental Results and Discussions}

Table~\ref{table:exp_results} presents the results of our experiments comparing different image encoding methods. We conducted experiments in various settings to obtain a comprehensive understanding of the methods: \\
(1) The ``Simple-plot" method serves as a straightforward baseline, where we plotted the ECG time series signals of 12 leads, selecting one ECG pause from each lead and putting them in a 4 $\times$ 3 layout. \\
(2) Using each encoding method individually, we formulated three baseline approaches referred to as ``MTF-only", ``GAF-only", and ``RP-only". \\
(3) Two encoding methods were randomly selected from the three and concatenated in different image channels. \\
(4) Instead of encoding each lead independently, we concatenated the 12 leads of one ECG pause into a single vector, which we then visualized using all three encoding methods. This approach is referred to as ``All-Concat".\\
(5) Finally, we gridded all three encoding methods in three image channels under both zero-shot and fine-tune settings, referred to as ``All-Grid (zero-shot)" and ``All-Grid (fine-tune)", respectively.

Table~\ref{table:exp_results} presents the experimental findings, which indicate that for single encoding comparison, RP encoding significantly outperforms both MTF and GAF encoding techniques. Moreover, the combination of GAF and RP encoding demonstrates superior performance compared to the other two combinations. Remarkably, the ``All-Grid (fine-tune)" method exhibits the best overall performance among all the baseline methods. A detailed analysis of the zero-shot and fine-tuning results shows that fine-tuning has a considerable impact on improving performance.

The ``All-Grid (fine-tune)" method utilizes a fine-tuning approach to improve the performance of the models by iteratively adjusting the parameters of the network. This method achieves the best overall performance by effectively leveraging the available data. The analysis of the zero-shot and fine-tuning results indicates that fine-tuning significantly enhances the performance of the models, highlighting the importance of optimizing the network parameters to improve the accuracy of the predictions.

\begin{table*}[htp]
\centering
\caption{Experimental results.}
\vspace{3pt}
\begin{center}
\begin{adjustbox}{width=0.75\linewidth}
\begin{tabular}{l|rrr|rrr|r}
\toprule 
\multirow{2}{*}{Method} & \multicolumn{3}{c}{Report Retrieval}  &\multicolumn{3}{c}{ Image Retrieval}  \\
& R@1 & R@5& R@10  &R@1 & R@5 & R@10  &RSUM \\
\midrule
Simple-plot &5.51 &17.16 &25.64 &4.98 &17.08 &26.05 &96.42 \\
\midrule
MTF-only &1.42 &5.67 &10.72 &1.97 &6.07 &10.80 &36.65  \\
GAF-only &2.69 &10.53 &17.04 &3.22 &11.01 &17.60 &62.09  \\
RP-only &5.27 &16.84 &25.68 &5.63 &17.20 &26.44 &97.06 \\
\midrule
MTF+GAF &4.02 &14.03 &21.91 &4.10 &14.42 &21.91 &80.39 \\
GAF+RP &6.30 &21.04 &30.36 &6.93 &20.69 &30.65 &115.97 \\
PR+MTF &5.12 &17.26 &25.84 &4.96 &17.10 &26.56 &96.84 \\
\midrule
All-Concat &1.58 &7.25 &12.77 &1.73 &7.33 &13.71 &4.37 \\
All-Grid (zero-shot) &0.21 &1.06 &1.91 &0.43 &1.06 &1.91  &6.58 \\
All-Grid (fine-tune) &\textbf{7.88} &\textbf{24.51} &\textbf{34.04} &\textbf{8.27} &\textbf{23.96} &\textbf{34.91} &\textbf{133.57}  \\
\bottomrule
\end{tabular}
\end{adjustbox}
\end{center}
\label{table:exp_results}
\vspace{-10pt}
\end{table*}

\subsection{Ablation Study}

In any modeling exercise, a vast array of parameters and settings can be adjusted to optimize performance. However, it is not always clear which of these factors has the most significant impact on the final output. In order to gain a better understanding of the inner workings of our model and the effect that each individual parameter has on its performance, we conducted a series of ablation studies.

To gain insight into the impact of batch size on the training and testing of our model, we conducted an ablation study. In this study, we systematically varied the training and testing batch sizes, and the results are presented in Table~\ref{table:ablation_batch}. Surprisingly, we found that a smaller training batch size led to better performance. This observation may seem counterintuitive, as larger batch sizes are typically favored in deep learning to accelerate training. However, our results suggest that a smaller training batch size may help the model converge more quickly and reduce overfitting. In addition, we observed that a smaller testing batch size also contributed to improved performance, when the training batch size was kept the same. This finding highlights the importance of matching the testing batch size to the training batch size, to ensure that the model is evaluated on a representative sample of data. By carefully selecting the appropriate training and testing batch sizes, we can optimize our model and achieve better results.

\begin{table}[htp]
\centering
\caption{Ablation study on learning batch.}
\vspace{3pt}
\begin{center}
\begin{adjustbox}{width=0.99\linewidth}
\begin{tabular}{l|ccc|ccc|c}
\toprule 
\multirow{2}{*}{Batch} & \multicolumn{3}{c}{Report Retrieval}  &\multicolumn{3}{c}{ Image Retrieval}  \\
& R@1 & R@5& R@10  &R@1 & R@5 & R@10  &RSUM \\
\midrule
Training 8 + Testing 32 &\textbf{7.88} &\textbf{24.51} &\textbf{34.04} &\textbf{8.27} &\textbf{23.96} &\textbf{34.91} &\textbf{133.57} \\
Training 16 + Testing 32 &7.88 &22.54 &32.23 &7.32 &21.91 &33.41 &125.29 \\
Training 32 + Testing 32 &7.01 &20.88 &32.23 &6.78 &21.04 &32.47 &120.41  \\
\bottomrule
\end{tabular}
\end{adjustbox}
\end{center}
\label{table:ablation_batch}
\vspace{-10pt}
\end{table}

In addition, we conducted an ablation study on the selection of the visual encoder. There are two choices of visual encoder selection: ViT-base and ViT-large. ViT-base has 12 transformer layers, about 85 million parameters, and is trained on images resized to 224x224 pixels. It is a relatively smaller model and is suitable for smaller datasets or where memory or computational resources are limited. ViT-large has 24 transformer layers, about 307 million parameters, and is trained on images resized to 384x384 pixels. It is a more complex and larger model, which typically results in better performance on large-scale datasets. 

The results of the ablation study are summarized in Table~\ref{table:ablation_vit}. Notably, we observed that for both the All-Grid and All-Concat settings, ViT-base outperformed ViT-large in terms of classification accuracy. These results are surprising, as ViT-large has more parameters and has been shown to perform better than ViT-base on pretraining tasks. However, we posit that the reason for this discrepancy lies in the size of the ECG image dataset. Specifically, since the dataset is relatively small, the fine-tuned embeddings of the ViT-base can more quickly adapt to the unique features of ECG images. In contrast, ViT-large contains more parameters and may require a larger dataset for effective fine-tuning. These findings have important implications for the use of ViT in medical image analysis. While ViT has shown great promise in a variety of visual recognition tasks, it is important to carefully consider its performance when applied to medical imaging datasets. Our results suggest that ViT-base may be a better choice for small medical imaging datasets, while ViT-large may be more effective for larger datasets. Additionally, our study highlights the importance of conducting careful evaluations of visual encoders in the medical imaging domain to ensure that they perform well on the specific imaging modality in question.

\begin{table}[htp]
\centering
\caption{Ablation study on vision encoder.}
\vspace{3pt}
\begin{center}
\begin{adjustbox}{width=0.99\linewidth}
\begin{tabular}{l|rrr|rrr|r}
\toprule 
\multirow{2}{*}{Vision Encoder} & \multicolumn{3}{c}{Report Retrieval}  &\multicolumn{3}{c}{Image Retrieval}  \\
& R@1 & R@5& R@10  &R@1 & R@5 & R@10  &RSUM \\
\midrule
All-concat (ViT-base) &2.52 &8.91 &13.79 &2.52 &7.25 &12.77 &47.76 \\
All-concat (ViT-large) &1.58 &7.25 &12.77 &1.73 &7.33 &13.71 &44.37  \\
\midrule
All-Grid (ViT-base) &\textbf{7.88} &\textbf{24.51} &\textbf{34.04} &\textbf{8.27} &\textbf{23.96} &\textbf{34.91} &\textbf{133.57} \\
All-Grid (ViT-large) &3.47 &12.32 &21.43 &4.73 &14.26 &22.14 &78.35 \\
\bottomrule
\end{tabular}
\end{adjustbox}
\end{center}
\label{table:ablation_vit}
\vspace{-10pt}
\end{table}

\section{Discussion and Conclusion}

Based on the experiments above, we have observed the strong potential of transforming ECG time series signals into images. Furthermore, by incorporating state-of-the-art advancements in vision-language learning, additional advantages can be gained from these encoded images. Our proposed model suggests that jointly learning the encoded ECG images and doctor's reports can yield improved representations. These representations hold promise for various clinical applications, including retrieving relevant previous diagnosis reports from a database. This support and reference can greatly assist doctors, leading to enhanced patient treatment outcomes.
Given the critical nature of healthcare, enhancing patient care remains of utmost importance. Introducing the proposed model into clinical applications has the potential to reshape the healthcare landscape and significantly influence patient outcomes. Therefore, we believe that our proposed model holds substantial practical value in the realm of clinical applications, offering significant advantages for patients, doctors, and the broader healthcare ecosystem.

\paragraph{Limitations}

While our study has illuminated the potential of MTF, GAF, and RP methods for ECG data analysis, it's important to acknowledge that other encoding techniques might also yield favorable outcomes. Furthermore, the dataset size used in our study might not comprehensively cover the variability and intricacy of ECG signals. Additionally, the accuracy of the doctor's report, a component of multimodal learning, could pose limitations due to inter-observer variability, potentially impacting the quality of learned representations. Moreover, variables like patient demographics, medical history, and comorbidities were not considered, suggesting an avenue for future exploration to enhance the generalizability of our findings. Hence, more extensive research, utilizing larger and more diverse datasets, encompassing various analysis techniques, and accounting for confounding factors, is crucial to fully delve into the potential of our approach.

\section*{Acknowledgements}

The research is partially supported by the DARPA ADAPTER program, and partially supported by the Allegheny Health Network and Mario Lemieux Center for Innovation and Research in EP.

\clearpage

\bibliography{qiu23}

\clearpage
\appendix

\section{Model Parameters}
\vspace{-10pt}
\begin{table}[htp]
\centering
\caption{Model parameters in the experiments.}
\vspace{5pt}
\begin{center}
\begin{adjustbox}{width=0.99\linewidth}
\begin{tabular}{c|c}
\toprule 
Parameters & Value \\
\midrule
alpha & 0.4 \\
k\_test & 128 \\
weight decay & 0.05 \\
queue size & 57600 \\
\bottomrule
\end{tabular}
\hspace{0.5cm}
\begin{tabular}{c|c}
\toprule 
ViT-base & Value \\
\midrule
train batch size & 16 \\
test batch size &16 \\
ViT layer & 4 \\
init lr & 1e-5 \\
\bottomrule
\end{tabular}
\hspace{0.5cm}
\begin{tabular}{c|c}
\toprule 
ViT-large & Value \\
\midrule
train batch size & 16 \\
test batch size &16 \\
ViT layer & 10 \\
init lr & 5e-6 \\
\bottomrule
\end{tabular}
\end{adjustbox}
\end{center}
\label{table:hyperparam}
\vspace{-15pt}
\end{table}

\section{Encoding Methods}\label{sec:appendix-encoding}

\subsubsection{Markov Transition Field (MTF)}

Markov Transition Field (MTF) is a method of transforming time series data, such as ECG signals, into visual representations. MTF works by calculating transition probabilities between adjacent data points in a time series, and then using these probabilities to generate a matrix of color-coded pixels. Each pixel in the matrix represents a unique transition probability, with darker colors indicating higher probabilities and lighter colors indicating lower probabilities. This matrix can be thought of as an image that encapsulates the key features of the original time series, making it easier for researchers and clinicians to analyze and interpret ECG signals. The development of the Markov Transition Field (MTF) draws inspiration from prior research on the interrelationship between time series and complex networks  \citep{Campanharo2011DualityBT,Zheng2014EEGbasedEC,Wang2014EncodingTS}. 
In essence, the MTF methodology involves constructing a Markov matrix based on quantile bins, which are derived through the discretization of the time series data. The dynamic transition probability of the time series is then encoded into a quasi-Gramian matrix, facilitating further analysis and interpretation of the underlying complex system.

In order to preserve time-domain information, the proposed method leverages Markov transfer probability to represent the dynamics of a given time series $X$. Specifically, the $Q$ quantile bins are identified, and each data point $x_i$ is assigned to its corresponding bin $q_j (j\in [1,Q])$. The resulting weighted adjacency matrix $W$, constructed using a first-order Markov chain model along the time axis, reflects the transition probabilities among the quantile bins. The frequency with which a data point in quantile bin $q_j$ is followed by a point in bin $q_i$ determines the value of the corresponding entry $w_{i,j}$ in $W$. Although $W$ represents the Markov transition matrix after normalization by $\sum_{j} { w_{ij}} = 1$, it is insensitive to the distribution of $X$ and the temporal dependencies between time steps $t_i$, resulting in a loss of information. To address this issue, the Markov Transition Field (MTF) $M$ is defined as follows:

\begin{equation}\scriptsize
\begin{bmatrix}
w_{ij|{x_1 \in q_i},{x_1 \in q_j}}&w_{ij|{x_1 \in q_i},{x_2 \in q_j}}    & \cdots\ &w_{ij|{x_1 \in q_i},{x_n \in q_j}}\\
w_{ij|{x_2 \in q_i},{x_1 \in q_j}}&w_{ij|{x_2 \in q_i},{x_2 \in q_j}}    & \cdots\ & w_{ij|{x_2 \in q_i},{x_n \in q_j}}\\
\vdots &\vdots   & \ddots  & \vdots  \\
w_{ij|{x_n \in q_i},{x_1 \in q_j}} &w_{ij|{x_n \in q_i},{x_2 \in q_j}}  & \cdots\ & w_{ij|{x_n \in q_i},{x_n \in q_j}}\\
\end{bmatrix}
\end{equation}

It involves building a $Q \times Q$ Markov transition matrix $W$ by dividing the time series data into $Q$ quantile bins, where $q_i$ and $q_j (q\in [1,Q])$ represent the quantile bins that contain the data at time stamps $i$ and $j$ along the temporal axis. The MTF matrix $M$ encodes the transition probabilities of the time series by spreading out the transition probability values from matrix $W$ along the magnitude axis to $M$ while taking into consideration the temporal positions. At each pixel $M_{ij}$, the probability of transitioning from the quantile at time step $i$ to the quantile at time step $j$ is assigned. In this way, the MTF matrix $M$ captures the multi-span transition probabilities of the time series. The entry $M_{i,j||i-j|=k }$ in $M$ represents the transition probability between points with a time interval of $k$, where $M_{i,j||j-i}=1$ represents the transition process along the time axis with a skip step. The main diagonal $M_{ii}$ in $M$ is a special case when $k = 0$ and captures the probability of transitioning from each quantile to itself, i.e., the self-transition probability, at time step $i$.

\subsubsection{Gramian Angular Field (GAF)}

Gramian Angular Field (GAF) \citep{Wang2014EncodingTS} is another method for transforming ECG time series signals into visual representations. GAF generates a matrix of cosine and sine values based on the pairwise differences between the original data points in the time series. This matrix is then transformed into an image, where each pixel corresponds to a particular combination of cosine and sine values. Similar to MTF, the resulting image captures important features of the original ECG signal, such as patterns and trends, which can aid in the interpretation and analysis of the data. The advantage of GAF over MTF is that it preserves the phase information of the original time series, which can be important in some applications, such as detecting arrhythmias.

The Gramian Angular Field (GAF) \citep{Wang2014EncodingTS} method represents time series data in a polar coordinate system instead of using the traditional Cartesian coordinates. In the Gramian matrix of GAF, each element corresponds to the cosine of the summation of angles.
The rescaled time series $\tilde{X}$ of $n$ real-valued observations are transformed to fall within the range of [$-1,1$] or [$0,1$] using the formula:
\begin{equation}\scriptsize
\tilde{x}^i_{-1}=\frac{(x_i -max(X) +(x_i -min(X))}{max(X)-min(X)}
\end{equation}
\begin{equation}\scriptsize
or \quad \tilde{x}^i_{0}=\frac{x_i -min(X)}{max(X)-min(X)}
\end{equation}
Then, by encoding the value as the angular cosine and the time stamp as the radius, we represent the rescaled time series $\tilde{X}$ in polar coordinates as follows:
\begin{equation}\scriptsize
\phi=arccos(\tilde{x}_i), \quad -1 \le \tilde{x}_i \le 1, \quad \tilde{x}_i \in \tilde{X}, \quad r=\frac{t_i}{N}, \quad t_i \in N
\end{equation}
Here, $t_i$ is the time stamp, and $N$ is a constant factor that regulates the span of the polar coordinate system. This encoding technique is a novel way to visualize time series data, where the values transform among different angular positions on the spanning circles as time passes, resembling water rippling. The encoding map is bijective, and it preserves absolute temporal relations, unlike Cartesian coordinates. The angular cosine function is monotonic for $\phi \in [0,\pi]$, producing a unique result in the polar coordinate system with a one-to-one inverse map.

Rescaled data in different intervals have different angular bounds. [0,1] corresponds to the cosine function in $ [0, \pi /2 ]$, while cosine values in the interval [$-1$,1] fall into the angular bounds $[0,\pi]$. They can provide different information granularity in the Gramian Angular Field for classification tasks, and the Gramian Angular Difference Field (GADF) of [0,1] rescaled data has an accurate inverse map. 

We utilize the angular perspective of the polar coordinate system to examine temporal correlations between different time intervals by calculating the trigonometric sum/difference between each point. Specifically, we define the Gramian Summation Angular Field (GASF) and Gramian Difference Angular Field (GADF) as follows:
\begin{equation}\scriptsize
\begin{aligned}
GASF=[cos({\phi}_i + {\phi}_j)]=\tilde{X}' \cdot \tilde{X} - \sqrt{I-\tilde{X}^2}' \cdot \sqrt{I-\tilde{X}^2}
\end{aligned}
\end{equation}
\begin{equation}\scriptsize
\begin{aligned}
GADF=[sin({\phi}_i - {\phi}_j)]=\sqrt{I-\tilde{X}^2}' \cdot \tilde{X} - \tilde{X}' \cdot \sqrt{I-\tilde{X}^2}
\end{aligned}
\end{equation}
Here, $I$ is the unit row vector $[1,1,...,1]$. After transforming the time series into the polar coordinate system, we treat each time step as a 1-D metric space. Defining the inner product as follows:
\begin{equation}\scriptsize
\begin{aligned}
<x,y >_1= x\cdot y - \sqrt{1 - x^2 } \cdot \sqrt{1 - y^2}
\end{aligned}
\end{equation}
\begin{equation}\scriptsize
\begin{aligned}
< x,y >_2=\sqrt{1 - x^2 } \cdot y - x \cdot \sqrt{1 - y^2}
\end{aligned}
\end{equation}
The two types of Gramian Angular Fields (GAFs) are actually quasi-Gramian matrices $[< \tilde{x_1}, \tilde{x_1} >]$.

The Gramian Angular Fields (GAFs) offer multiple benefits. First, they enable the retention of temporal relationships, as the position's movement from the top-left to the bottom-right corresponds to the increase in time. The GAFs incorporate temporal correlations since $G_{i,j||i-j|=k}$ symbolizes the relative correlation due to the superimposition/difference of directions concerning time interval $k$. The main diagonal $G_{i,i}$ is a special case for $k=0$, containing the original angular/value information.

\subsubsection{Recurrence Plot (RP)}

Recurrence Plot (RP) \citep{Eckmann1987RecurrencePO} is a non-linear time series analysis technique that can also be applied to transform ECG time series signals into visual representations. RP generates a square matrix that reflects the similarity between all pairs of data points in the time series. The matrix is constructed by measuring the distance between each pair of data points and comparing them to a predefined threshold value. If the distance between two points is below the threshold, the corresponding matrix element is set to 1, otherwise, it is set to 0. This results in a binary matrix that can be visualized as an image, where dark pixels represent recurrent patterns in the time series. RP has been shown to be effective in capturing complex patterns in ECG signals, such as P-waves and QRS complexes, which are important for the accurate diagnosis of cardiovascular diseases. 

Given a time series $(x_1, \ldots, x_n)$, we can extract trajectories from it as follows:
\begin{equation}\scriptsize
\begin{aligned}
\vec{x}i = (x_i, x{i + \tau}, \ldots, x_{i + (m - 1)\tau}), \quad
\forall i \in {1, \ldots, n - (m - 1)\tau }
\end{aligned}
\end{equation}
Here, $m$ denotes the dimension of the trajectories, and $\tau$ is the time delay. Once we have extracted the trajectories, we can create a recurrence plot, denoted by $R$, which is essentially the pairwise distance between the trajectories. Formally, we define $R_{i, j}$ as:
\begin{equation}\scriptsize
\begin{aligned}
R_{i, j} = \Theta(\varepsilon - | \vec{x}_i - \vec{x}_j |), \quad
\forall i,j \in {1, \ldots, n - (m - 1)\tau }
\end{aligned}
\end{equation}
Here, $\Theta$ is the Heaviside step function, and $\varepsilon$ is the threshold. The recurrence plot helps us visualize the structure and patterns of the time series by preserving the temporal dependencies and revealing the relative correlations between the extracted trajectories.

\section{Loss Objectives}

There are three objectives during learning, including Image-Text Contrastive (ITC) Loss, Image-Text Matching  (ITM) Loss, and Mask Language Modeling (MLM) Loss.  An overview of each loss is provided in the subsequent sections. More details can be found in the Appendix due to the page limit.

\paragraph{Image-Text Contrastive Loss (ITC)}  The Image-Text Contrastive Loss (ITC) loss has been shown to be highly effective in improving vision and language understanding in a range of applications, including image captioning, visual question answering, and multimodal retrieval~\citep{clip, ALBEF}. To compute the ITC loss, we follow the approach proposed by~\citet{ALBEF}, which introduces a momentum encoder to generate features and creates soft labels from the momentum encoder to serve as training targets. The soft labels help account for the potential positive samples in the negative pairs and improve the quality of the learned representations. 
Our model learns a similarity function represented by 
$
s=g_v(\Vec{v}_\mathrm{cls})^\top g_w(\Vec{w}_\mathrm{cls}), 
$
which aims to increase the similarity scores for matching image-text pairs. Here, $g_v$ and $g_w$ refer to linear transformations that convert the \texttt{[CLS]} embeddings into lower-dimensional, normalized (256-d) representations. Following the MoCo approach~\citep{moco}, we use two queues to store the most recent $M$ image-text representations obtained from the momentum unimodal encoders. The features obtained from the momentum encoders are normalized and denoted by $g'_v(\Vec{v}'_\mathrm{cls})$ and $g'_w(\Vec{w}'_\mathrm{cls})$. To calculate the similarity score between an image-text pair and a text-image pair, we define $s(I,T) = g_v(\Vec{v}_\mathrm{cls})^\top g'_w(\Vec{w}'_\mathrm{cls})$ and $s(T,I) = g_w(\Vec{w}_\mathrm{cls})^\top g'_v(\Vec{v}'_\mathrm{cls})$, respectively.

We use the softmax-normalized image-to-text and text-to-image similarity to calculate each image and text. This is represented by the equations below, where $\tau$ is a temperature parameter that can be learned:
\begin{equation}\scriptsize
\label{eqn:sim}
p_m^\mathrm{i2t}(I) = \frac{\exp (s(I,T_m) / \tau)}{\sum_{m=1}^M \exp (s(I,T_m)/ \tau)},
\end{equation}
\begin{equation}\scriptsize
p_m^\mathrm{t2i}(T) = \frac{\exp (s(T,I_m)/ \tau)}{\sum_{m=1}^M \exp (s(T,I_m)/ \tau)}
\end{equation}
We represent the ground-truth one-hot similarity as $\Vec{y}^\mathrm{i2t}(I)$ and $\Vec{y}^\mathrm{t2i}(T)$, where negative pairs have a probability of 0, and the positive pair has a probability of 1. The image-text contrastive loss is defined as the cross-entropy $\mathrm{H}$ between $\Vec{p}$ and $\Vec{y}$, which is shown in the following equation:
\begin{equation}\scriptsize
\label{eqn:itc}
\mathcal{L}\mathrm{itc} = \frac{1}{2} \mathbb{E}{(I,T)\sim D} \big[ \mathrm{H}(\Vec{y}^\mathrm{i2t}(I),\Vec{p}^\mathrm{i2t}(I)) + \mathrm{H}(\Vec{y}^\mathrm{t2i}(T),\Vec{p}^\mathrm{t2i}(T)) \big]
\end{equation}

\paragraph{Image-Text Matching Loss (ITM)} The Image-Text Matching Loss (ITM) is responsible for activating the image-grounded text encoder, with the goal of learning a multimodal representation that captures the detailed alignment between visual and linguistic information. ITM is a binary classification task where the model predicts whether an image-text pair is positive (matched) or negative (unmatched) based on its multimodal feature. The ITM head, which is a linear layer, is used to make this prediction.

To obtain the joint representation of the image-text pair, we use the output embedding of the \texttt{[CLS]} token from the multimodal encoder, and then append a fully-connected (FC) layer followed by softmax to predict a two-class probability $p^\mathrm{itm}$. The ITM loss is defined as:
\begin{equation}\scriptsize
\label{eqn:itm}
\mathcal{L}\mathrm{itm} = \mathbb{E}{(I,T)\sim D} \mathrm{H} (\Vec{y}^\textrm{itm}, \Vec{p}^\textrm{itm}(I,T))
\end{equation}
where $\Vec{y}^\textrm{itm}$ is a 2-dimensional one-hot vector representing the ground-truth label.
To improve the selection of negative pairs, we employ a strategy called hard negative mining, as proposed by~\citet{ALBEF}. This strategy involves selecting negative pairs that have a higher contrastive similarity within a batch, as they are more informative and can improve the learning process.

\paragraph{Mask Language Modeling Loss (MLM)} The Mask Language Modeling Loss (MLM) is used to predict masked words using both the image and contextual text. In this loss, we randomly mask out input tokens with a probability of 15\% and replace them with the special token \texttt{[MASK]}, with 10\% random tokens, 10\% unchanged, and 80\% \texttt{[MASK]} replacements following the BERT approach. The predicted probability of a masked token is denoted by $\Vec{p}^\textrm{msk}(I,\hat{T})$, where $\hat{T}$ represents the masked text. The cross-entropy loss is used to minimize the difference between the predicted and ground-truth distributions, which is expressed as follows:
\begin{equation}\scriptsize
\label{eqn:mlm}
\mathcal{L}\mathrm{mlm} = \mathbb{E}{(I,\hat{T})\sim D} \mathrm{H} (\Vec{y}^\textrm{msk}, \Vec{p}^\textrm{msk}(I,\hat{T}))
\end{equation}
$\Vec{y}^\textrm{msk}$ represents a one-hot vocabulary distribution and the ground-truth token has a probability of 1.

\section{More Related Works}

\paragraph{Machine Learning in ECG}

With the development of machine learning and deep learning, many works have studied the application of using advanced models in ECG. 
\citet{Alfaras2019AFM} proposed a fully automatic and fast ECG arrhythmia classifier based on a simple brain-inspired machine learning approach known as Echo State Networks. 
\citet{Mishra2020ECGPR} converted ECG paper records into a 1-D signal and generated an accurate diagnosis of heart-related problems using deep learning. \citet{Peimankar2020DENSECGAD} combined CNN and long LSTM model to detect the onset, peak, and offset of different heartbeat waveforms such as the P-wave, QRS complex, T-wave, and No wave (NW). 
\citet{Aziz2021ECGbasedMA} exploited two-event related moving-averages (TERMA) and fractional-Fourier-transform (FrFT) algorithms. 
\citet{Somani2021DeepLA} proposed a review focusing on orienting the clinician towards fundamental tenets of deep learning, state-of-the-art prior to its use for ECG analysis, and current applications of deep learning on ECGs. 
\citet{Kim2022MachineLB} proposed a ML model for real-time classification of atrial fibrillation (AF) between Paroxysmal atrial fibrillation (PAF) and Non-paroxysmal atrial fibrillation (Non-PAF). 
\citet{Adedinsewo2022DigitizingPB} evaluated how well the AI-ECG model output obtained using digitized paper ECGs agreed with the predictions from the native digital ECGs for the detection of low ejection fraction. \citet{Ayano2022InterpretableML} summarized the achievements in ECG signal interpretation using interpretable machine learning techniques. 
\citet{Qiu2023Trans} proposed an approach for cardiovascular disease diagnosis and automatic ECG diagnosis report generation.
\citet{Zhu2022GeoECGDA,Qiu2022CardiacDD,Qiu2022OptimalTB} proposed a physiologically-inspired data augmentation method to improve performance and increase the robustness of heart disease detection based on ECG signals.

\paragraph{Transform Time Series Signals into Images}

Encoding  time series data as different types of images have been explored by many studies in different areas \citep{Wang2014EncodingTS}. \citet{Hatami2017ClassificationOT} used Recurrence Plots (RP) to transform time series into 2D texture images. 
\citet{Kavasidis2017Brain2ImageCB} converted brain signals into images. 
\citet{MartnezArellano2019ToolWC} proposed an approach for tool wear classification based on signal imaging.  
\citet{Barra2020DeepLA} encoded financial time series into images to predict the future trend of the U.S. market.  
\citet{Qin2020ImagingAF} encoded time series of sensor data as images to retain necessary features for human activity recognition. 
\citet{Bi2021TourismDF} transformed time series into images to improve the accuracy of tourism demand forecasting. 
\citet{Yuan2021EncodingTG} developed a new image encoding technique based on time-series segmentation (TS) to transform acceleration (A), velocity (V), and displacement (D) ground motion records into a three-channel AVD image of the ground motion event. 
\citet{Sayed2023FromTT} transformed multivariate time-series data into images for better encoding and extracting relevant features for non-intrusive occupancy detection. 


\end{document}